\documentstyle[preprint,aps,epsfig]{revtex}

\def\be{\begin{equation}}
\def\ee{\end{equation}}
\def\ep{\epsilon}
\def\bea{\begin{eqnarray}}
\def\eea{\end{eqnarray}}

\begin{document}
\draft

\title{Dielectric behaviour of graded spherical cells \\
 with an intrinsic dispersion}
\author{Y. T. C. Ko$^{1,2}$, J. P. Huang$^{1,3}$, K. W. Yu$^1$}
\address{$^1$Department of Physics, The Chinese University of Hong Kong,
 Shatin, NT, Hong Kong}
\address{$^2$Trinity College, University of Cambridge,
 Cambridge CB2 1TQ, United Kingdom}
\address{$^3$Max Planck Institute for Polymer Research, Ackermannweg 10, 
 55128 Mainz, Germany}

\maketitle

\begin{abstract}
The dielectric properties of single-shell spherical cells with an 
intrinsic dielectric dispersion has been investigated. By means of the 
dielectric dispersion spectral representation (DDSR) for the 
Clausius-Mossotti (CM) factor, we express the dispersion strengths as 
well as the characteristic frequencies of the CM factor analytically in 
terms of the parameters of the cell model. These analytic expressions 
enable us to assess the influence of various model parameters on the 
electrokinetics of cells. Various interesting behaviours have been 
reported. We extend our considerations to a more realistic cell model 
with a graded core, which can have spatial gradients in the conductivity 
and/or permittivity. To this end, we address the effects of a graded
profile in a small-gradient expansion in the framework of DDSR.
\end{abstract}
\vskip 5mm

\section{Introduction}

The interaction of the polarization of biological cells with the applied 
fields has resulted in a wide range of practical applications from 
manipulation, trapping to separation of biological cells~\cite{Jones}, 
and even nanotechnology~\cite{Hughes}. When a biological cell in medium is
exposed to an applied electric field, there is an accumulation of charge 
at the interfaces and hence a dipole moment is induced in the cell. 
The strength of the polarization depends on the frequency of the applied 
field as well as on the permittivities and conductivities of cells and medium. 
The situation becomes more complicated when we consider 
structured particles because biological cells are usually modeled as 
conductive spheres (cytosol) with a thin insulating outer shell (membrane),
assuming the shell is an isotropic, non-dispersive dielectric with
conductive losses. In this case, there are additional 
frequency-dependent changes in the polarization.

The Clausius-Mossotti (CM) factor determines the polarization of a 
biological particle in a surrounding medium, and is a measure of the 
dielectric contrast between the particle and medium. The CM factor is 
important in biophysical research because it is closely related to the 
alternating current (ac) electrokinetic behaviors of biological cells, 
namely, dielectrophoresis~\cite{Pohl}, electrorotation~\cite{Arnold}, 
electro-orientation~\cite{Saito}, electrofusion~\cite{Zimmermann-0}, 
as well as electrodeformation~\cite{VL-11}. Any change in the cell's
properties such as the mobile charges, or particle shape as well as the 
variation of medium conductivity or medium permittivity will change the 
CM factor, which is in turn reflected in the ac electrokinetic spectra. 
These spectra show characteristic frequency-dependent changes amongst other
complicated features.

Moreover, the conductivities and permittivities can have characteristic 
frequency dependencies due to the presence of mobile charges in 
membrane. Thus the constancy of these quantities is only an 
approximation and these quantities do change with frequency, giving rise 
to additional dispersions. In this work, we aim to establish a 
dielectric dispersion spectral representation (DDSR) for the 
single-shell spherical cell model with an intrinsic dielectric 
dispersion in the cytosol. 
The DDSR was pioneered by Maxwell \cite{Maxwell} in 1891 in the context 
of interfacial polarization. When two media are put in contact (thus 
forming an interface) and an electric field is applied, polarization 
charge is induced at the interface due to the dielectric contrast 
between the two media. Although Maxwell considered a two phase system in 
which one phase is insulating, it can be readily generalized to a more 
general case when both media have complex dielectric permittivities. 

The DDSR was subsequently extended to spherical particles by Lei {\it et 
al.}~\cite{Lei} and further elaborated by Gao {\it et al.}~\cite{Gao} 
for cell models without shells, the single-shell model has been widely 
used to mimic a living biological cell as a homogeneous, nondispersive 
spherical particle surrounded by a thin shell corresponding to the 
plasma membrane.
The DDSR enables us to express the CM factor analytically in terms of a 
series of sub-dispersions, each of which with analytic expressions for 
the dispersion strengths and their corresponding characteristic 
frequencies expressed in terms of the various parameters of the cell 
model~\cite{Lei,Gao}. Thus this representation enables us to assess in 
detail the influence of the various model parameters, including 
structural, material, as well as dynamic properties of cells, without 
the need to analyze the full dielectric dispersion spectrum.

The paper is organized as follows. In the next section, we review the 
dielectric dispersion spectral representation (DDSR) for the CM factor 
of an unshelled spherical cell model~\cite{Lei} to establish notations. 
We express the dispersion strength as well as the characteristic 
frequency of the CM factor analytically in terms of the parameters of 
the cell model. Then an intrinsic dielectric dispersion is included in 
the cell~\cite{Gao}. In Section III, we analyze the single-shell model 
with a dispersive core and a non-dispersive, insulating shell. We apply 
DDSR to the CM factor to obtain the analytic expressions for the dispersion
strengths and characteristic frequencies. These expressions enable us to 
assess the influence of various model parameters on the electrokinetics 
of cells. In Section IV, we examine the influence of the individual 
parameters, such as the conductivities of the external medium and the 
cytosol on the dispersion spectra. Various interesting behaviours will 
be obtained. In Section V, we extend our considerations to a graded 
core, namely, the core can have spatial gradients in the conductivity 
and/or permittivity. We address the effects of a graded profile in a 
small-gradient expansion in the general framework of DDSR. Discussion and
conclusion will be given in Section VI.

\section{Dielectric dispersion spectral representation}

In this section, we review the dielectric dispersion spectral
representation for the CM (Clausius-Mossotti) factor of an
unshelled spherical cell model~\cite{Lei}. The dipole moment $p$
of a single sphere in uniform electric field~\cite{Jackson} \be p
= {\ep_e\ U\ D^3 \over 8}\ E_0 \ee where $\ep_e$ is the
permittivity of the external medium, $D$ is the diameter of the
particle and $E_0$ is the electric field strength. $U$ is the CM
factor due to the dielectric discontinuity and follows the
equation \be U = {\ep_i - \ep_e \over \ep_i +2\ep_e} \ee where
$\ep_i$ is the permittivity of the particle. In AC applied fields,
we replace the permittivities with their complex counterparts:
\bea
\ep_i \to {\ep_i^*} = \ep_i + {\sigma_i \over i\omega}, \\
\ep_e \to {\ep_e^*} = \ep_e + {\sigma_e \over i\omega},
\eea
where $i=\sqrt{-1}$, $\sigma_i$ and $\sigma_e$ are conductivities of the particle and of the external
medium respectively. Then
\be
U \to {U^*} = {{\ep_i^*} - {\ep_e^*} \over {\ep_i^*} +2{\ep_e^*}}
\ee
This gives the dielectric relaxation of a single spherical particle
\be
{U^*} = U + {\Delta \ep \over 1 + i\omega / \omega_c}
\ee
with the characteristic frequency $\omega_c$ and dispersion strength
$\Delta \ep$:
\bea
\omega_c &=& {\sigma_i + 2\sigma_e \over \ep_i + 2\ep_e}, \\
\Delta \ep &=& {\sigma_i - \sigma_e \over \sigma_i +2\sigma_e}
 - {\ep_i - \ep_e \over \ep_i +2\ep_e}.
\eea It is related to the Maxwell-Wagner structure relaxation
$\omega_c = 10^4\,$ s$^{-1} \cdots 10^9\,$s$^{-1}$.

The angular velocity $\Omega$ of electrorotation is
\be
\Omega = - {\ep_e E_0^2\over 2 \eta} Im\ {U^*},
\ee
where $\eta$ is the coefficient of viscosity.
Note that  $Im\ {U^*} < 0$ gives co-field rotation while $Im\ {U^*} > 0$ gives
anti-field rotation.

Then, when an intrinsic dielectric dispersion is included in the
cell~\cite{Gao}, we again replace the permittivities with the
appropriate complex counterparts: \bea \ep_i^* &=& \ep_i +
 {\Delta\ep_i \over 1+i\omega/\omega_c} + {\sigma_i\over i\omega},\\
\ep_e^* &=& \ep_e + {\sigma_e \over i\omega}.
\eea

The corresponding complex CM factor $U_{\rm{int}}{}^*$ can then be
expressed in the dispersion terms as
\begin{equation}
U_{\rm{int}}{}^*=U_{\rm{int}}+\sum_{n=1}^{2}\frac{\Delta\ep_n}{1+i\omega/\omega_n},
\label{dispersion}
\end{equation}
where $U_{\rm{int}}=(\ep_i-\ep_e)/(\ep_i+2\ep_e)$, $\Delta\ep_n$s are the dispersion strengths and
$\omega_n$s are the characteristic frequencies.

To solve for the dispersion strengths and the characteristic frequencies, assume the summation term
in Eq.~(\ref{dispersion}) is of the form
\bea
U_{\rm{int}}{}^*-U_{\rm{int}} &=& \frac{P_0+P_1w}{1+R_1w+R_2w^2}\\
&=& \frac{P_0+P_1w}{(1+w/\omega_1)(1+w/\omega_2)}
\eea
where $w=i\omega$ and, $P$s and $R$s are constants in terms of the model parameters.

For the characteristic frequencies, solve the following quadratic
equation \be 1+R_1w+R_2w^2=0 \ee and the $\omega_n$s are {\em
minus} the solutions to the equation. They come out to be, in
terms of the model parameters, 
\bea \omega_1 &=&
\frac{1}{2(2\ep_e+\ep_i)}[2\sigma_e+\sigma_i+(\Delta\ep_i+2\ep_e+\ep_i)\omega_c+
\sqrt{\Gamma}], \\
\omega_2 &=&
\frac{1}{2(2\ep_e+\ep_i)}[2\sigma_e+\sigma_i+(\Delta\ep_i+2\ep_e+\ep_i)\omega_c-
\sqrt{\Gamma}], 
\eea 
where 
\begin{equation} \Gamma =
-4(2\ep_e+\ep_i)(2\sigma_e+\sigma_i)\omega_c+[2\sigma_e+\sigma_i+(\Delta\ep_i+2\ep_e+\ep_i)
\omega_c]^2. 
\end{equation}

For the dispersion strengths, performing partial fraction can express the summation term in the form
of the summation term in Eq.~(\ref{dispersion}). The dispersion strengths turn out to be, in terms of
model parameters and characteristic frequencies,
\bea
\Delta\ep_1 &=&
\frac{3(-\ep_i\sigma_e\omega_1+\ep_e\sigma_i\omega_1+\ep_i\sigma_e\omega_c-
\ep_e\sigma_i\omega_c+\Delta\ep_i\ep_e\omega_1\omega_c)}
{(2\ep_e+\ep_i)^2\omega_1(\omega_1-\omega_2)},\\
\Delta\ep_2 &=&
\frac{3(\ep_i\sigma_e\omega_2-\ep_e\sigma_i\omega_2-\ep_i\sigma_e\omega_c+
\ep_e\sigma_i\omega_c-\Delta\ep_i\ep_e\omega_2\omega_c)}
{(2\ep_e+\ep_i)^2\omega_2(\omega_1-\omega_2)}. 
\eea

It is worth remarking that, two dispersion terms
appear in Eq.~(\ref{dispersion}): the first term (i.e. when
$n=1$) is due to the phase difference between the cell and the
medium, and the second term (i.e. when $n=2$) is due to
the presence of the intrinsic dispersion inside the cell.

This is a special case of the model mentioned in the following section. It is interesting to compare
that this model, with no shell, but the same core as the next model, has two dispersion strengths
(and the same number of characteristic frequencies), while the next model, with shell, has three
dispersion strengths (and the same number of characteristic frequencies).

Similar work was done by Foster {\it et al.}~\cite{Foster}. For
the case of nondispersive particle and medium, our solutions are
indeed equivalent to those of Foster {\it et al.} However, for the case
of dispersive particle and nondispersive medium, we quoted the
exact analytic solutions while Foster {\it et al.} only presented the
approximate solutions obtained by expanding the exact solutions
using Taylor's expansion (cf. Section b of Ref.~\cite{Foster}).

\section{Single-shell spherical cell model, with a dispersive core}

The CM (Clausius-Mossotti) factor of an isotropic model with a
non-dispersive homogeneous core has been
investigated~\cite{Sheng,Zimmermann}. Here we would like to establish
the DDSR (dielectric dispersion spectral representation) of an isotropic
model with a {\em dispersive} homogeneous core covered with a
non-dispersive, insulating membrane~\cite{Gimsa-2}.

The idea of DDSR is to mathematically extract the analytic
expressions of the dispersion strengths and the corresponding
characteristic frequencies from the CM factor. The CM factor for a
single-shell spherical cell with isotropic, lossless dielectric
membrane is~\cite{Sheng,Zimmermann} \be\label{U_{iso}} U_{iso} = \frac
{(2\ep_m+\ep_i)(\ep_m-\ep_e)R_e^3 +
(\ep_i-\ep_m)(2\ep_m+\ep_e)R_i^3}
{(2\ep_m+\ep_i)(2\ep_e+\ep_m)R_e^3+2(\ep_i-\ep_m)(\ep_m-\ep_e)R_i^3},
\ee where $\ep$ is permittivity and $R$ the radius; the subscripts
$e$, $m$ and $i$ correspond to the external medium, the membrane
and the cytosol respectively.

For adaptation to our concerned model, the real constants $\ep_e$,
$\ep_m$ and $\ep_i$ are replaced by the complex counterparts \bea
\label{i}\ep_i^* &=& \ep_i + \frac {\Delta\ep_i}
{1+i\omega/\omega_d} + \frac {\sigma_i}
{i\omega}\\
\label{m}\ep_m^* &=& \ep_m + \frac {\sigma_m} {i\omega}\\
\label{e}\ep_e^* &=& \ep_e + \frac {\sigma_e} {i\omega}.
\eea
The complex $\ep_i^*$ contains the dispersive term ($\frac {\Delta\ep_i} {1+i\omega/\omega_d}$) to
account
for the intrinsic dispersive nature of the cytosol, while both the membrane and the external medium
are non-dispersive.

The CM factor becomes complex and can be written as \be\label{sum}
U_{dis}^* = U_{iso} + \sum_{t=1}^{3} \frac {\Delta\ep_t}{1+
i\omega/\omega_t} \ee where $\Delta\ep_t$ is the dispersion
strengths, and $\omega_t$ is the characteristic frequencies.

$\Delta\ep_t$ and $\omega_t$ can be solved easily using
\textit{Mathematica}. Assume the summation part to be of the form
\bea
U_{dis}^* - U_{iso} &=& \frac {B_0+B_1w+B_2w^2} {1+A_1w+A_2w^2+A_3w^3} \\
&=& \frac
{B_0+B_1w+B_2w^2}{(1+w/\omega_1)(1+w/\omega_2)(1+w/\omega_3)}.
\eea where $w = i\omega$ and the $A$s and $B$s are constants in
terms of the parameters of the model. Performing partial fraction
can express this term in the form of the summation in
Eq.~(\ref{sum}).

To solve for $\omega_t$, solve the cubic equation
\be
1+A_1w+A_2w^2+Aw^3=0
\ee
$\omega_t$s are {\em minus} the solutions to this equation.

$\Delta\ep_1$ in terms of the constants $B$s and $\omega_t$ is
\be
\Delta\ep_1 = \frac
{(B_0+\omega_1(-B_1+B_2\omega_1))\omega_2\omega_3}{(\omega_1-\omega_2)(\omega_1-\omega_3)}
\ee
The rest of the $\Delta\ep_t$s follow by cyclic permutation of the variables, namely, $1\rightarrow2,
2\rightarrow3 \mbox{ and } 3\rightarrow1$.

\section{The influence of individual parameters}

This model depends on the thickness of the membrane, the
permittivities and conductivities of three different regions (i.e.
the cytosol, the membrane and the external medium) and the
properties of the cytosol dispersion. Using \textit{Mathematica}
these parameters can be varied individually. Each time only one
parameter is varied, while the rest are kept at the values in
Table~\ref{values}. These variations show interesting behaviours.

As shown in the figures, there are three sub-dispersions:
$\Delta\ep_1$ being the co-field peak related to the cytosol,
$\Delta\ep_2$ being the anti-field peak related to the membrane
and $\Delta\ep_3$ being the anti-field peak related to the
intrinsic dispersion of the cytosol.

In Fig.~1, the high-frequency co-field dispersion strength
$\Delta\ep_1$ remains relatively constant from $\sigma_e =
1\times10^{-5}\,$S/m to about $\sigma_e = 0.01\,$S/m and then
decreases with increasing $\sigma_e$, due to a significant
reduction in the conductivity contrast between the cytosol and the
external medium. Its corresponding characteristic frequency
$\omega_1$ also remains relatively constant in the mentioned range
and then increases with increasing $\sigma_e$. The anti-field
dispersion strengths $\Delta\ep_2$ and $\Delta\ep_3$ and their
corresponding characteristic frequencies $\omega_2$ and $\omega_3$
show more interesting behaviours. $\Delta\ep_2$ and $\Delta\ep_3$
swap at between $\sigma_e = 0.00018\,$S/m and $\sigma_e = 0.00019\,$S/m,
while $\omega_2$ and $\omega_3$ show level-repulsion, i.e. their
values gain closer, being closest at the same value of $\sigma_e$
as when the swapping occurs, and then their values move apart
again. This phenomenon is very common in many physical systems and
is frequently observed in atomic physics. These interesting
phenomena are evidences that both $\Delta\ep_2$ and $\Delta\ep_3$
are real (as opposed to virtual solutions arising from inaccurate
calculations) and are common in varying many of the parameters, as
shown below.

In Fig.~2, increasing $\sigma_i$ causes $\Delta\ep_1$ to increase
from negative (anti-field) to positive (co-field) and then remain
constant, $\Delta\ep_2$ to decrease to a constant value and
$\Delta\ep_3$ to remain constant throughout. $\omega_1$ increases
monotonically while $\omega_2$ and $\omega_3$ remains roughly
constant.

In Fig.~3, varying $\omega_d$ has negligible effect on
$\Delta\ep_1$ and thus also $\omega_1$. In fact, $\Delta\ep_2$ and
$\Delta\ep_3$ are not very much affected if not for the swapping
occurring at about $\omega_d = 30000$\,rad/s. Their
corresponding characteristic frequencies also show
level-repulsion, as in previous cases, with the closest point also
at about $\omega_d = 30000$\,rad/s.

In Fig.~4, $\Delta\ep_1$ and $\Delta\ep_3$ (and also their
corresponding characteristic frequencies $\omega_1$ and
$\omega_3$) show negligible variations. Both $\Delta\ep_2$ and
$\omega_2$ remain relatively constant before increasing. They
being the only affected ones because they are, as well as the
concerned parameter $\sigma_m$, related to the membrane.

Compared with the isotropic mobile charge model with a non-dispersive homogeneous core previously
investigated, it is interesting that the variations of different permittivities and conductivities
show remarkably similar results, with the most noticeable difference that the swapping and the
level-repulsion did not occur in the previous model.

Fig.~5 shows the real and
imaginary parts of the CM factor  against the field frequency for several
values of the medium conductivity, in an attempt  to illustrate the results in Fig.~1. In this figure, two dispersions are observed. In fact, as shown in Fig.~1, the third dispersion strength is small enough to be neglected, and hence the third dispersion in Fig.~5 cannot be shown, as expected. Similarly, we are able to adjust the other parameters respectively, like the cytosolic conductivity, circular frequency of cytosol dispersion and external conductivity, in order to illustrate the results in Figs.~2$\sim$4. However, all of them should show a  framework similar to Fig.~5, and hence are omitted.

\section{Small-gradient expansion}

After investigating models with homogeneous cores, it is natural
for us to proceed to investigate models with non-homogeneous
cores. Here we choose to investigate a model that consists of a
dispersive core with graded dielectric profile, and a
non-dispersive membrane. We consider a graded permittivity profile
\be \ep_i(r) = \ep_i + ar + O[a]^2, \ \ \ 0 \le r \le R_i, \ee
where $a$ is a gradient, which has the unit as permittivity per
unit length. We start from Eq.(\ref{U_{iso}}) and replace the
cytosolic permittivity by an equivalent permittivity
$\bar{\ep}_i(r)$~\cite{Yu} \be \bar{\ep}_i(r) = \ep_i +
\frac{3}{4}a R_i + O[a]^2. \label{ave} \ee As shown below,
$\bar{\ep}_i(r)$ can formally be calculated using the
small-gradient expansion of the differential effective dipole
approximation (DEDA)~\cite{Yu,Huang}.

After the substitution, we can expand the CM factor using Taylor's
expansion, up to the second order:
\be
U(a)=U(0)+aU'(0)+O[a]^2.
\ee
The first term is the same as $U_{iso}$, while the second term is the
correction due to the graded profile.

We now extract the DDSR of the second term as usual, by replacing
the permittivities by their complex counterparts as in
Eqs.(\ref{i})--(\ref{e}), using the following substitution \be X^*
= \frac{X}{i\omega} \ee

Using these substitutions, we can see that the dielectric profile is
\bea
\ep_i^*(r) = \ep_i + \frac {\sigma_i + ar} {i\omega} +O[a]^2
\eea

Assume the second term has the form
\bea
aU'(0) &=& \frac {C_0+C_1w+C_2w^2+C_3w^3}{1+D_1w+D_2w^2+D_3w^3+D_4w^4}\\
&=& \frac {C_0+C_1w+C_2w^2+C_3w^3}{((1+w/\omega_1)(1+w/\omega_2))^2}
\eea
where $w=i\omega$ and $C$s and $D$s are constants in terms of the parameters of the model.

Although the denominator is a quartic equation, there are only two distinct solutions for the
characteristic frequency. This is due to the differentiation performed in the Taylor's expansion,
causing each frequency to split into a repeated root.

By partial fraction $aU'(0)$ takes the form
\be
\sum_{q=1}^2 \frac{\Delta\ep_q}{1+ \frac{i\omega}{\omega_q}} + \sum_{q=1}^2
\frac{\Delta^2\ep_q}{(1+ \frac{i\omega}{\omega_q})^2}
\ee
To avoid confusion, it should be remarked that $\Delta^2\ep_q$ does not equal the square of
$\Delta\ep_q$.

To solve for $\omega_q$, solve the quartic equation
\be
(1+D_1w+D_2w^2)^2=0.
\ee
$\omega_q$s are {\em minus} the solutions to this equation.

$\Delta\ep_1$ and $\Delta^2\ep_1$ in terms of the constants $C$s and the characteristic frequencies
$\omega_q$s are
\bea
\Delta\ep_1 &=& \frac
{\omega_1\omega_2^2(2C_0-C_1(\omega_1+\omega_2)+\omega_1(C_3\omega_1(\omega_1-3\omega_2)+2C_2\omega_2))}
{(\omega_1-\omega_2)^3},\\
\Delta^2\ep_1 &=& \frac
{(C_0-\omega_1(C_1+\omega_1(-C_2+C_3\omega_1)))\omega_2^2}{(\omega_1-\omega_2)^2}.
\eea $\Delta\ep_2$ and $\Delta^2\ep_2$ can be obtained by
replacing $\omega_1$ with $\omega_2$ and $\omega_2$ with
$\omega_1$.

We are now in a position to show how to find $\bar{\ep}_i(r)$ from
DEDA~\cite{Yu,Huang}. For the dipole factor of a graded spherical
particle, the following differential equation
holds~\cite{Yu,Huang} \be \frac{db}{dr} = -\frac
{1}{3r\ep_e\ep_i(r)}
 [(1+2b)\ep_e-(1-b)\ep_i(r)][(1+2b)\ep_e+2(1-b)\ep_i(r)]
\ee where $b$ is the dipole factor, $r$ is the radius, and
$\ep_i(r)$ is the dielectric profile.

Since
\be
b(r) = \frac {\bar{\ep}_i(r)-\ep_e}{\bar{\ep}_i(r)+2\ep_e}
\ee
solving for $b(r)$ is equivalent to solving for $\bar{\ep}_i(r)$.

Since we are doing a small-gradient expansion, $b(r)$ can be expressed as
\be
b(r) = b_0 + b_1 +O[a]^2
\ee
where
\be
b_0 = \frac{\ep_i - \ep_e}{\ep_i +2\ep_e}
\ee
and $b_1$ can be solved from the differential equation
\be
\frac{db_1}{dr} =
-\frac{3[(\ep_i+2\ep_e)^2b_1-3\ep_ear]}{r(\ep_i+2\ep_e)^2}.
\ee
The solution reads
\be
b_1 = \frac {9a\ep_e}{4(\ep_i+2\ep_e)^2}
\ee
using the initial condition at $r=0$.

After putting all the pieces together, $\bar{\ep}_i(r)$ comes out as in
Eq. (\ref{ave}).

Using the parameters as shown in Table~\ref{values} and $a =
0.025/R_i$ (where $R_i = R_e-d$ and is the internal radius), we
have done some numerical calculations.  This value of $a$
corresponds to a change of 10\% over the internal radius.  The
results are shown in Table~\ref{results}. These are corrections to
the calculations in the previous session due to a small gradient.
The characteristic frequencies remain the same as in the previous
isotropic electrostatic model (except now each is a repeated
root), while the dispersion strengths are smaller than those in
the previous model by one to two order of magnitude. This shows
that our small-gradient expansion is valid.

\section{Discussion and Conclusion}

Here a few comments are in order.
In view of our recent success in the DDSR of single-shell spherical cell
model, we are prepared to illustrate the DDSR in various different
situations. We would like to extend DDSR to cell suspensions of higher
concentration. At a higher concentration, we expect mutual interactions 
among cells and the dielectric behaviors can change significantly. We 
may extend DDSR to polydisperse cells, because the cells may have 
different sizes and/or permittivities. The polydispersity can have 
nontrivial impact on their dielectric behaviors. Eventually we have to 
overcome the analytic continuation, and analyze the dispersion spectrum 
of the full anisotropic mobile charge model and the graded cell model.

Regarding the applicability of the Clausius-Mossotti approach, one can solve the electrostatic problem first, and then extend to complex permittivities accordingly, as pointed out by Jones~\cite{Jones}.  As a matter of fact, there are already theories of e.g. Maxwell and Wagner, and
 Rayleigh for heterogeneous dielectrics~\cite{Scaife}. In this regard, it is of value to compare these theories with the present approach.

   In the present paper, we have discussed isolated particles in the dilute limit. In fact, for higher volume fractions, we can use the effective-medium
theories instead~\cite{Gao}, like Maxwell-Garnett approximation or Effective Medium Approximation.

Throughout the paper, the cells under consideration exist in the form of a spherical shape. In this connection, we may include the 
shape effect as well. More precisely, we can extend the graded spheroidal cell 
model of Huang {\it et al.}~\cite{Huang} to include an intrinsic 
dispersion in the core. This is a nontrivial extension and we believe 
the non-spherical shape will have significant impact on the dispersion 
spectrum. We can consider the following agenda: (1) homogeneous spheroidal
cell with intrinsic dispersion, without shell~\cite{Gao}; (2)
graded spheroidal cell with intrinsic dispersion, without shell.
Also, item~2 will be studied in the small-gradient expansion.

In view of the present interesting results,  the corresponding experiment is suggested to be done. In doing so, one may use coated colloids having a graded core.

In summary, we have considered a single-shell model with an 
inhomogeneous graded cytosol. Realistic cells must be inhomogeneous due 
to the compartment in the interior of cells. In such a model, the 
cytosol can have a conductivity profile which varies along the radius of 
the cell. A small conductivity-gradient expansion for the DDSR of 
single-shell graded cell model has been done, based on the differential 
effective dipole approximation~\cite{Yu,Huang}. We have assessed the 
effects of a conductivity gradient in the cytosol on the dispersion 
spectrum.

\section*{Acknowledgments}

This work was supported in part by the RGC Earmarked Grant under project
number CUHK 403303 and in part by the Direct Grant for Research.
Y.T.C.K. wishes to thank the Jardine Foundation Scholarship (2002) for
supporting her education at the University of Cambridge.

\begin{table}
\begin{tabular}{lll}
\hline
Parameters&Symbols&Numerical Values\\
\hline
Cell radius & $R_e$ & 9.5\,$\mu$m\\
Membrane thickness & $d$ & 8\,nm\\
External permittivity & $\ep_e$ & 80$\ep_0$\\
External conductivity & $\sigma_e$ & 1\,mS/m\\
Cytosolic permittivity & $\ep_i$ & 120$\ep_0$\\
Cytosolic conductivity & $\sigma_i$ & 0.25\,S/m\\
Cytosolic dielectric increment & $\Delta\ep_i$ & 800$\ep_0$\\
Membrane permittivity & $\ep_m$ & 7.23$\ep_0$\\
Membrane conductivity & $\sigma_m$ & $4\times10^{-7}\,$S/m\\
Circular frequency of cytosol dispersion & $\omega_d$ &  $10^4$\,rad/s\\
\hline
\end{tabular}
\caption{Parameters used for model calculations}
\label{values}
\end{table}

\begin{table}
\begin{tabular}{lll}
\hline
Solutions&Symbols&Absolute numerical Values\\
\hline
Characteristic frequencies & $\omega_1$ & $1.03\times10^8$rad/s\\
& $\omega_2$ & $3.18\times10^4$rad/s\\
& $\omega_3$ & $1.00\times10^4$rad/s\\
\\
Dielectric dispersion strengths & $\Delta\ep_1$ & $0.0621$\\
& $\Delta\ep_2$ & $-0.00133$\\
& $\Delta\ep_3$ & $1.42\times10^{-8}$\\
& $\Delta^{2}\ep_1$ & $-0.0613$\\
& $\Delta^{2}\ep_2$ & $0.000539$\\
& $\Delta^{2}\ep_3$ & $2.02\times10^{-12}$\\
\hline
\end{tabular}
\caption{Results from small-gradient model calculations}
\label{results}
\end{table}

\newpage

\begin{figure}[ht]
\caption{The dispersion strengths ($\Delta
\epsilon_1$~$\sim$~$\Delta \epsilon_3$) and the characteristic
frequencies ($\omega_1$~$\sim$~$\omega_3$) as a function of the
conductivity of the external medium $\sigma_e.$}
\end{figure}

\begin{figure}[ht]
\caption{Same as Fig.~1, but as a function of the conductivity of
the cytosol $\sigma_i .$ Typical $\sigma_i$ values range from
$0.2$ to $1\,$S/m.}
\end{figure}

\begin{figure}[ht]
\caption{Same as Fig.~1, but as a function of the circular
frequency of the cytosol dispersion $\omega_d .$}
\end{figure}

\begin{figure}[ht]
\caption{Same as Fig.~1, but as a function of the conductivity of
the membrane $\sigma_{m} .$}
\end{figure}

\begin{figure}[ht]
\caption{Real and imaginary parts of the CM factor as a function of the circular frequency of the external field. ${\rm Re}[\cdots]$  (${\rm Im}[\cdots]$) denotes the real (imaginary) part of $\cdots .$ }
\end{figure}

\newpage
\centerline{\epsfig{file=se.eps,width=440pt}} \centerline{Fig.1}

\newpage
\centerline{\epsfig{file=si.eps,width=440pt}} \centerline{Fig.2}

\newpage
\centerline{\epsfig{file=wd.eps,width=440pt}} \centerline{Fig.3}

\newpage
\centerline{\epsfig{file=sm.eps,width=440pt}} \centerline{Fig.4}

\newpage
\centerline{\epsfig{file=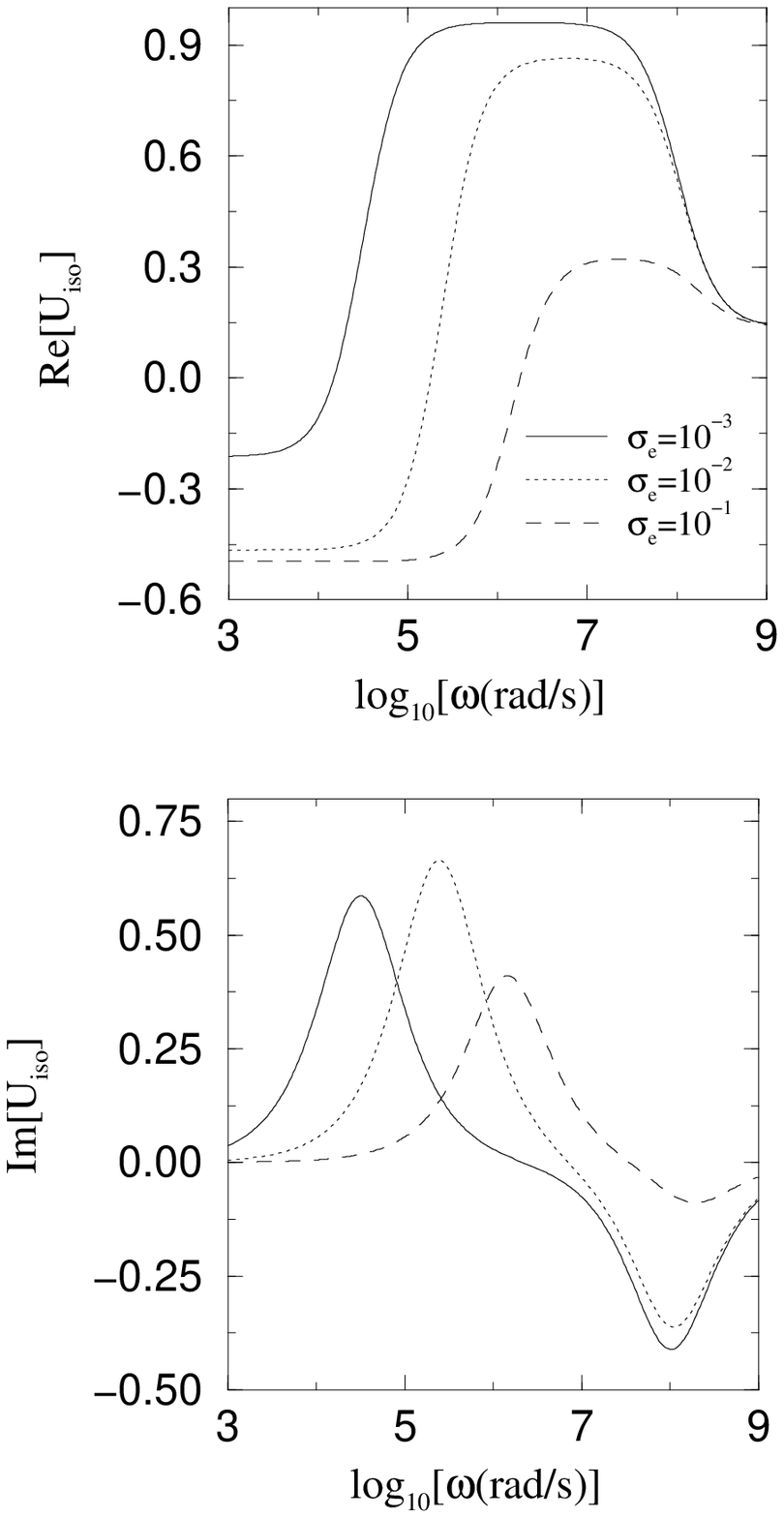,width=200pt}} \centerline{Fig.5}

\end{document}